\documentclass{article}

\usepackage{graphicx}

\newcommand{\x}{\overline{x}}
\newcommand{\y}{\overline{y}}
\newcommand{\ri}{{\rm i}}
\newcommand{\e}{{\rm i}\epsilon}
\newcommand{\nn}{\nonumber}
\newcommand{\eir}{\varepsilon}
\newlength{\slas}
\newcommand{\Sl}[1]{\ensuremath{\settowidth{\slas}{$#1$}
\hspace{0.5\slas}\settowidth{\slas}{$/$}\hspace{-0.5\slas}/
\settowidth{\slas}{$/$}\hspace{-0.5\slas}\settowidth{\slas}
{$#1$}\hspace{-0.5\slas}}#1}

\begin{document}

\begin{titlepage}

\begin{flushright}
IRB$-$TH$-$2/03
\end{flushright}
\vspace{1.cm}

\begin{center}
\large\bf
{\LARGE\bf Reduction method for dimensionally regulated
one-loop N-point Feynman integrals}\\[1cm]
\rm
{G. Duplan\v ci\' c \footnote{e-mail: gorand@thphys.irb.hr} and 
B. Ni\v zi\' c \footnote{e-mail: nizic@thphys.irb.hr}}\\[1cm]

{\em Theoretical Physics Division, Rudjer Bo\v{s}kovi\'{c} Institute, 
        P.O. Box 180, HR-10002 Zagreb, Croatia} \\[.5cm]
      
\end{center}
\normalsize

\vspace{1cm}

\begin{abstract}
We present a systematic method for reducing 
an arbitrary one$-$loop $N-$point massless
Feynman integral with generic 4$-$dimensional 
momenta to a set comprised of eight fundamental
scalar integrals: six box integrals in $D=6$, a triangle
integral in $D=4$, and a general two$-$point integral
in $D$ space time dimensions.
All the divergences present in the original integral
are contained in the general two$-$point integral
and associated coefficients.
The problem of vanishing of the kinematic determinants has
been solved in an elegant and transparent manner.
Being derived with no restrictions regarding the external
momenta, the method is completely general and applicable
for arbitrary kinematics. In particular, it applies to
the integrals in which the set of external momenta contains
subsets comprised of two or more collinear momenta,
which are unavoidable when calculating one$-$loop contributions
to the hard$-$scattering amplitude for exclusive hadronic
processes at large momentum transfer in PQCD.
The iterative structure makes it easy to implement the
formalism in an algebraic computer program.
\end{abstract}


\end{titlepage}

\section{Introduction}

Scattering processes have played a crucial role 
in establishing the fundamental interactions of nature.
They represent the most important source of information
on short-distance physics.
With increasing energy, multiparticle
events are becoming more and more dominant.
Thus, in testing various aspects of QCD, the high-energy scattering
processes, both exclusive and inclusive, in which the total
number of particles (partons) in the initial and final states
is $N\ge 5$, have
recently become increasingly important.

Owing to the well-known fact that the LO predictions in
perturbative QCD (PQCD) do not have much predictive power, the
inclusion of higher-order corrections is essential for
many reasons.
In general, higher-order corrections have a stabilizing 
effect,
reducing the dependence of the LO predictions on the
renormalization and factorization scales and the 
renormalization scheme.
Therefore, to achieve a complete confrontation between
theoretical predictions and experimental data, it is very
important to know the size of radiative corrections to the
LO predictions.

Obtaining radiative corrections requires the evaluation
of one-loop integrals arising from the Feynman diagram
approach.
With the increasing complexity of the process under
consideration, the calculation of radiative corrections
becomes more and more tedious.
Therefore, it is extremely useful to have an algorithmic
procedure for these calculations, which is computerizable
and leads to results which can be easily and safely evaluated
numerically.

The case of Feynman integrals with massless internal 
lines is of special interest,
because one often deals with either really massless particles
(gluons) or particles whose masses can be neglected in
high$-$energy processes (quarks).
Owing to the fact that these integrals contain 
IR divergences (both soft and collinear), they need 
to be evaluated in an arbitrary number of space-time
dimensions.
As it is well known, in calculating Feynman diagrams 
mainly three difficulties
arise: tensor decomposition of integrals, reduction of scalar
integrals to several basic scalar integrals and the evaluation
of a set of basic scalar integrals.

Considerable progress has recently been made in developing
efficient approaches for calculating one$-$loop Feynman
integrals with a large number ($N\ge 5$) of external lines
\cite{old,davy,dixon,bern,tarasov,tarasov1,binoth,
binoth1,glover,denner}.
Various approaches have been proposed for reducing the
dimensionally regulated ($N\ge 5$)$-$point tensor integrals 
to a linear combination of $N-$ and lower$-$point scalar
integrals multiplied by tensor structures made from the
metric tensor $g^{\mu\nu}$ and external momenta
\cite{old,davy,tarasov,binoth,denner}.
It has also been shown that the general $(N > 5)-$point
scalar one$-$loop integral can recursively be represented
as a linear combination of $(N-1)$$-$point
integrals provided the external momenta are kept in four
dimensions \cite{dixon,bern,tarasov,tarasov1,binoth,binoth1}.
Consequently, all scalar integrals occurring in the
computation of an arbitrary one$-$loop $(N\ge 5)$$-$point
integral can be reduced to
a sum over a set of basic scalar box $(N=4)$ integrals
with rational coefficients depending on the external momenta
and the dimensionality of space$-$time.
Despite, the considerable progress, the developed methods still cannot be
applied to all cases of practical interest. 
The problem is related to 
vanishing of various relevant kinematic determinants.

As far as the calculation of one-loop $(N > 5)$-point
massless integrals is concerned,
the most complete and systematic method is presented
in \cite{binoth}. It does not, however, apply to all 
cases of practical interest. Namely, being obtained for the
non-exceptional external momenta it cannot be, for example, applied
to the integrals in which the set of external momenta
contains subsets comprised of two or three collinear
on-shell momenta. 
The integrals of this type arise when
performing the leading-twist NLO analysis of hadronic
exclusive processes at large$-$momentum
transfer in PQCD.

With no restrictions regarding the external kinematics,
in this paper we formulate an efficient, systematic and
completely general method for reducing an arbitrary one$-$loop
$N-$point massless integral to a set of basic integrals.
Although the method is presented for massless case, the generalization on
massive case is straightforward. 
The main difference between the massive and massless cases manifests itself in
the basic set of integrals, which in former case is far more complex.
Among the one$-$loop Faynman integrals there exist both massive and massless
integrals for which the existing reduction methods break down.
The massless integrals belonging to this category are of more practical interest
at the moment, so in this paper we concentrate on massless case. 


The paper is organized as follows.
Section 2 is devoted to introducing notation and to
some preliminary considerations.
In Sec. 3, for the sake of completenes, we briefly review a tensor 
decomposition method for $N-$point tensor integrals which was 
originally obtained in
Ref. \cite{davy}. 
In Sec. 4 we present a procedure
for reducing one$-$loop $N$$-$point massless scalar integrals
with generic $4$$-$dimensional external momenta to a fundamental
set of integrals. 
Since the method is closely related to the one given in \cite{tarasov,tarasov1},
similarities and differences between the two are pointed out.
Being derived with no
restrictions to the external momenta, the method is completely
general and applicable for arbitrary kinematics.
Section 5 contains considerations regarding the
fundamental set of integrals which is comprised of eight integrals.
Section 6 is devoted to some concluding remarks.
In the Appendix A we give explicit expressions
for the relevant basic massless box integrals
in $D=6$ space$-$time dimensions. These integrals constitute
a subset of the fundamental set of scalar integrals.
As an illustration of the tensor decomposition and scalar reduction methods,
in the Appendix B we evaluate an one$-$loop 6$-$point Feynman diagram
with massless internal lines, contributing to the NLO hard$-$scattering
amplitude for $\gamma \, \gamma \rightarrow \pi^+ \, \pi^-$ exclusive reaction
at large momentum transfer in PQCD.

\section{Definitions and general properties}

In order to obtain one$-$loop radiative coorections
to physical processes in massless gauge theory,
the integrals of the following type are required:
\begin{equation}
I^N_{\mu _1\cdots \mu_P}(D;\{p_i\})\equiv
(\mu ^2)^{2-D/2}\int
\frac{{\rm d}^D l}{(2\pi)^D}
\frac{l_{\mu_1}\cdots l_{\mu_P}}{A_1A_2 \cdots
A_N}~\cdot
\label{f1}
\end{equation}
This is a rank $P$ tensor one$-$loop $N$$-$point Feynman 
integral with massless internal lines in $D$$-$di\-men\-si\-o\-nal
space$-$time,
where $p_i$, $(i=1,2,...,N)$ are the external momenta,
$l$ is the loop momentum, and $\mu$ is the usual
dimensional regularization scale.

\begin{figure}
\begin{center}
\includegraphics[width=10cm]{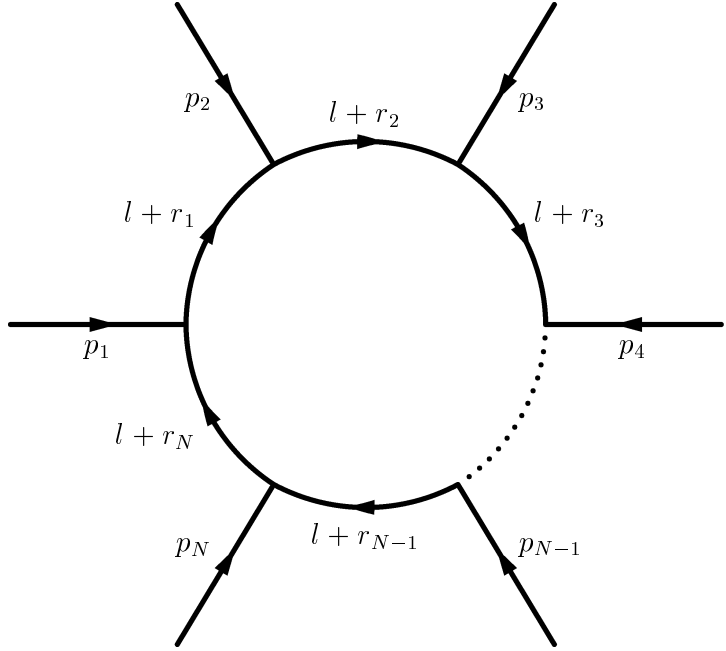}
\end{center}
\caption{\label{fig1}{\em One-loop N-point diagram}}
\end{figure}

The Feynman diagram with $N$ external lines,
which corresponds to the above integral, is 
shown in Fig. \ref{fig1}. 
For the momentum assignments as shown, i.e.
with all external momenta taken to be
incoming, the massless propagators have the form
\begin{equation}
A_i\equiv (l+r_i)^2+\e\qquad i=1,\cdots, N~,
\label{f2}
\end{equation}
where the momenta $r_i$ are given by $r_i=p_i+r_{i-1}$
for $i$ from 1 to $N$, and $r_0=r_N$.
The quantity $\e$($\epsilon >0$) represents
an infinitesimal imaginary part,
it ensures causality and after the integration
determines the correct sign of the imaginary
part of the logarithms and dilogarithms.
It is customary to choose the loop momentum in a such a way
that one of the momenta $r_i$ vanishes. However, for general 
considerations, such a choice is not convenient, since by
doing so, one loses the useful symmetry of the integral with
respect to the indices $1,\cdots ,N$. 

The corresponding scalar integral is
\begin{equation}
I^N_0(D;\{p_i\})\equiv
(\mu ^2)^{2-D/2}\int
\frac{{\rm d}^D l}{(2\pi)^D}
\frac{1}{A_1A_2 \cdots
A_N}~.
\label{f3}
\end{equation}
If $P+D-2N \geq 0$, 
the integral (\ref{f1}) is UV divergent.
In addition to UV divergence, the integral can 
contain IR divergence.
There are two types of IR divergence: collinear and soft.
A Feynman diagram with massless particles
contains a soft singularity if it contains an internal 
gluon line attached to two external quark lines which are 
on mass$-$shell.
On the other hand,
a diagram contains a collinear singularity
if it contains an internal gluon line attached to 
an external quark line which is on mass $-$shell.
Therefore, a diagram containing a soft singularity 
at the same time contains two collinear singularities, i.e.
soft and collinear singularities overlap.

When evaluating Feynman diagrams, one ought to regularize all divergences.
Making use of the dimensional regularization method, one can
simultaneously regularize UV and IR
divergences,
which makes the dimensional regularization method optimal for
the case of massless field theories.

The tensor integral (\ref{f1}) is, as it is seen, 
invariant under the permutations of the propagators
$A_i$, and is symmetric with respect to the Lorentz indices 
$\mu_i$. Lorentz covariance allows
the decomposition of the tensor integral (\ref{f1}) in the form
of a linear decomposition consisting of the momenta $p_i$ and the
metric tensor $g_{\mu \nu}$.

\section{Decomposition of tensor integrals}

Various approaches have been proposed for reducing the
dimensionally regulated $N-$point tensor integrals to
a linear combination of $N-$ and lower$-$point scalar
integrals multiplied by tensor structures made from the
metric tensor $g^{\mu\nu}$ and external momenta.
In this section we briefly review the derivation of the tensor 
reduction formula originally obtained in Ref. \cite{davy}. 

For the purpose of the following discussion,
let us consider the tensor integral
\begin{equation}
I^N_{\mu _1\cdots \mu_P}(D;\{{\nu}_i\})\equiv
(\mu ^2)^{2-D/2}\int
\frac{{\rm d}^D l}{(2\pi)^D}
\frac{l_{\mu_1}\cdots l_{\mu_P}}{A_1^{{\nu}_1}
A_2^{{\nu}_2} \cdots
A_N^{{\nu}_N}}~,\label{f4}
\end{equation}
and the corresponding scalar integral
\begin{equation}
I_0^N(D;\{{\nu}_i\})\equiv
(\mu ^2)^{2-D/2}\int
\frac{{\rm d}^D l}{(2\pi)^D}
\frac{1}
{A_1^{{\nu}_1}
A_2^{{\nu}_2} \cdots
A_N^{{\nu}_N}}~\cdot \label{f5}
\end{equation}
The above integrals represent generalizations of the
integrals (\ref{f1}) and (\ref{f3}), in
that they contain arbitrary powers ${\nu}_i\in \bf N$
of the propagators in the integrand,
where $\{\nu_i \}$ is the shorthand notation for
($\nu_1, \cdots ,\nu_{N}$). Also, for notational simplicity,
the external momenta are omitted from the argument of the
integral.

The Feynman parameter representation 
of the tensor integral
$I^N_{\mu _1\cdots \mu_P}(D;\{ \nu_i\})$, given in (\ref{f4}),
which is valid for arbitrary values of
$N$, $P$, $r_i$ and $\nu_i(>0)$, for the values
of $D$ for which the remaining integral is finite,
and the $\Gamma-$function does not diverge is given by
\begin{eqnarray}
\lefteqn{I^N_{\mu _1\cdots \mu_P}(D;\{ \nu_i\})=
\frac
{\ri }{(4\pi)^2}
(4\pi \mu ^2)^{2-D/2}\!\! 
\sum_{k,
j_1,\cdots,j_{N}\geq 0\atop 2 k+{\scriptstyle
\Sigma} j_i=P} \!\!\left \{ [g]^k [r_1]^{j_1}\cdots 
[r_{N}]^{j_{N}}\right\}_{\mu _1\cdots \mu_P}}\nn \\
& &\times\,\frac{\Gamma\left( 
\sum\nolimits_i \nu_i-D/2-k\right)}{2^k
\,\left[\prod\nolimits_i \Gamma
 (\nu_i)\right]}(-1)^{\Sigma_i \nu_i+P-k}\int_0^1\! \left(
  \prod\nolimits_i {\rm d}y_i
y_i^{\nu_i+j_i-1}\right)\nn \\ 
& &\times \delta\left(
\sum\nolimits_{i=1}^{N}y_i-1\right)
\left[ \,-\sum_{i,j=1\atop i<j}^N y_i y_j 
\left( r_i-r_j\right)^2
-\e  \, \right]^{k+D/2-\Sigma_i
 \nu_i},\label{f12}
\end{eqnarray}
where $ \{ [g]^k [r_1]^{j_1}\cdots 
[r_{N}]^{j_{N}}\}_{\mu _1\cdots \mu_P}$
represents a symmetric (with respect to $\mu _1\cdots \mu_P$)
combination of tensors, each term of which is composed
of $k$ metric tensors and $j_i$ external momenta $r_i$.
Thus, for example,
\[
\left \{ g r_1\right\}_{\mu _1 \mu_2 \mu_3}=
g_{\mu_1 \mu_2} r_{1\mu_3}+g_{\mu_1 \mu_3} r_{1\mu_2}+
g_{\mu_2 \mu_3} r_{1\mu_1}.
\]
As for the integral representation of the corresponding
scalar integral (\ref{f5}) the result is of the form
\begin{eqnarray}
\lefteqn{I^N_0(D;\{ \nu_i\})=\frac{
\ri }{(4\pi)^2}
(4\pi \mu ^2)^{2-D/2}\frac{\Gamma\left( 
\sum\nolimits_{i=1}^{N} \nu_i-D/2\right)}{
\prod\nolimits_{i=1}^{N} \Gamma
 (\nu_i)}(-1)^{\Sigma_{i=1}^N \nu_i}}\nn \\ 
& &\times\,\int_0^1\! \left( \prod_{i=1}^{N}
 {\rm d}y_i 
y_i^{\nu_i-1}\right)
\delta\left(
\sum_{i=1}^{N}y_i-1\right)
\left[ \,-\sum_{i,j=1\atop i<j}^N y_i y_j 
\left( r_i-r_j\right)^2
-\e  \, \right]^{D/2-\Sigma_{i=1}^{N}
 \nu_i}\hspace{-1.5cm}.\label{f13}
\end{eqnarray}
Now, on the basis of (\ref{f13}),
(\ref{f12}) can be written in the form
\begin{eqnarray}
\lefteqn{I^N_{\mu _1\cdots \mu_P}(D;\{ \nu_i\})= 
\sum_{k,
j_1,\cdots,j_{N}\geq 0\atop 2 k+{\scriptstyle
 \Sigma} j_i=P} \left \{ [g]^k [r_1]^{j_1}\cdots 
 [r_{N}]^{j_{N}}\right\}_{\mu _1\cdots \mu_P} } \nn \\
& &
{}\times \frac{( 4 \pi \mu^2)^{P-k}}{(-2)^{k}}
\left[\prod_{i=1}^N \frac{\Gamma (\nu_i+j_i)}{
\Gamma(\nu_i)}\right] I^N_0(D+2(P-k);\{ 
\nu_i+j_i\}). \label{f15}
\end{eqnarray}
This is the desired decomposition, 
of the dimensionally
regulated $N-$point rank $P$ tensor integral.
It is originally obtained in Ref. \cite{davy}.
Based on (\ref{f15}), any dimensionally regulated
$N-$point tensor integral can be expressed as
a linear combination of $N-$point scalar 
integrals multiplied by tensor structures made from the
metric tensor $g^{\mu\nu}$ and external momenta.
Therefore, with the decomposition (\ref{f15}), the problem 
of calculating the tensor integrals has been reduced to the 
calculation of the general scalar integrals.

It should be pointed out that among the tensor reduction methods presented in
the literature one can find methods, e.g. \cite{binoth} which for 
$N\ge 5$ completely avoid the terms proportional to the metric tensor 
$g^{\mu\nu}$. Compared with the method expressed by (\ref{f15}),
these reduction procedures lead to decomposition containing a smaller number
of terms. The methods of this type are based on the assumption that for 
$N\ge 5$, one can find four linearly independent 4-vectors forming a basis 
of the 4-dimensional Minkowski space, in terms of which the metric 
tensor can then be expressed. This assumption is usually not realized when 
analyzing
the exclusive processes at large-momentum-transfer (hard scattering picture) in
PQCD. Thus, for example, in order to obtain the next-to-leading order
corrections to the hard-scattering amplitude for the proton-Compton scattering,
one has to evaluate one-loop $N=8$ diagrams. The set of external momenta
contains two subsets comprised of three collinear momenta (representing the
proton). The kinematics of the process is thus limited to the 3-dimensional
subspace. If this is the case, the best way of doing the tensor decomposition is
the one based on formula (\ref{f15}), 
regardless of the fact that for large $N$ the number of terms obtained can be
very large.

As is well known, the direct evaluation of the 
general scalar integral (\ref{f5}) (i.e. (\ref{f13}))
represents a non-trivial problem.
However, with the help of the recursion relations,
the problem can be significantly
simplified in the sense that the calculation of
the original scalar integral
can be reduced to the calculation of a certain number
of simpler fundamental (basic) intgerals.

\section{Recursion relations for scalar integrals}

Recursion relations for scalar integrals have been known for some time
\cite{dixon,bern,tarasov,tarasov1,binoth}. However, as it turns out, the 
existing set of relations that can be
found in the literature is not sufficient to perform the reduction procedure
completely, i.e. for all one-loop integrals appearing in practice. The
problem is related to vanishing of various kinematic determinants, it is
manifest for the cases corresponding to $N>6$, and it is especially acute when
evaluating one-loop Feynman integrals appearing in the NLO analysis of
large momentum transfer exclusive processes in PQCD.
As is well known, these processes are generally described in terms of Feynman
diagrams containing a large number of external massless lines. Thus, for example,
for nucleon Compton scattering is $N=8$. A large number of external lines implies
a large number of diagrams to be considered, as well as a very large number of
terms generated when performing the tensor decomposition using (\ref{f15}).
In view of the above, to
treat the Feynman integrals (diagrams) with a large number of external lines
the use of computers is unavoidable. This requires that the scalar reduction
procedure be generally applicable. It is therefore absolutely clear that any 
ambiguity or uncertainty present in
the scalar recursion relations constitutes a serious problem.
The method presented below makes it possible to perform the reduction completely
regardless of the kinematics of the process considered and the complexity of the
structure of the contributing diagrams.

For the reason of completeness and clearness of presentation and with the aim of
comparison with the already existing results, we now briefly present a few main
steps of the derivation of recursion relations. It should be pointed out that
the derivation essentially represents a variation of the derivation originally
given in \cite{tarasov}.

Recursion relations for scalar integrals are obtained with the help of the
integration-by-parts method \cite{tarasov,davy1,ibp}. 
Owing to translational invariance, the dimensionally
regulated integrals satisfy the following identity:
\begin{equation}
0\equiv
\int
\frac{{\rm d}^D l}{(2\pi)^D}\frac{\partial}{\partial
l^\mu}\left(
\frac{z_0 l^{\mu}+\sum\nolimits_{i=1}^{N}
z_i r_i^{\mu}}{A_1^{\nu_1} \cdots
A_N^{\nu_N}}\right)~,
\label{f16}
\end{equation}
where $z_i~(i=0\cdots N)$ are arbitary constants, while
$A_i$ are the propagators given by (\ref{f2}).
The identity (\ref{f16}) is a variation of the identity used in 
\cite{tarasov}, where it was assumed that $r_N=0$.
Performing the differentiation,
expressing scalar products in the numerator  
in terms of propagators $A_i$, choosing
$z_0=\sum_{i=1}^N z_i$, 
(which we assume in the following)
and taking into account the scalar integral (\ref{f5}),
the identity (\ref{f16}) leads to the relation  
\begin{eqnarray}
\lefteqn{\sum_{j=1}^{N}\left(\sum_{i=1}^{N} \left[
(r_j-r_i)^2
+2\e\right] z_i\right) \nu_j I_0^N(D;\{\nu_k+\delta_{k j}\})
}\nn \\
&=&\!\!\!\sum_{i, j=1}^{N}z_i \nu_j
I_0^N(D;\{\nu_k+\delta_{k j}-\delta_{k i}\})
-
(D-\sum_{j=1}^{N}\nu_j) z_0
I_0^N(D;\{\nu_k\}), \label{f20}
\end{eqnarray}
where $\delta_{ij}$ is the Kronecker delta symbol.
In arriving at (\ref{f20}), it has been understood that
\begin{equation}
I_0^N(D;\nu_1,\cdots,
\nu_{l-1}, 0, \nu_{l+1},\cdots, \nu_N) \equiv
I_0^{N-1}(D;\nu_1,\cdots, \nu_{l-1}, \nu_{l+1},\cdots, \nu_N). \label{f21}
\end{equation}
The relation (\ref{f20}) represents
the starting point for the derivation 
of the recursion relations for scalar integrals.

We have obtained
the fundamental set of recursion relations 
by choosing the arbitrary constants $z_i$
so as to satisfy the following system of linear equations:
\begin{equation}
\sum\nolimits_{i=1}^N 
(r_i-r_j)^2 z_i=C,\quad j=1,\ldots, N,\label{f25}
\end{equation} 
where $C$ is an arbitrary constant.
Introducing the notation
$r_{ij}=(r_i-r_j)^2$, 
the system (\ref{f25}) may be written in matrix notation as
\begin{equation}
\left( \begin{array}{ccccc}
0 & r_{12} & \cdots & r_{1N} \\
r_{12} & 0 & \cdots & r_{2N} \\
\vdots & \vdots & \ddots & \vdots \\
r_{1N} & r_{2N} & \cdots & 0
\end{array} \right) 
\left( \begin{array}{c}
z_1 \\ z_2 \\ \vdots \\ z_N
\end{array} \right)
=
\left( \begin{array}{c}
C \\ C \\ \vdots \\ C
\end{array} \right) . \label{f27}
\end{equation}

It should be pointed out that the expression of the type (\ref{f20}) and the system of
the type  (\ref{f27}), for the case of massive propagators ($A_i= (l+r_i)^2-m_i^2+\e$),
 see Ref. \cite{tarasov}, can simply
be obtained from the relation given above by making a change $r_{ij}\rightarrow r_{ij}-m_i^2-m_j^2$.
Consequently, considerations performed for the massive case \cite{tarasov, tarasov1} apply to the
massless case, and vice versa.

It should be mentioned that, in the existing literature, 
the constant $C$ used
to be chosen as a real number different from zero. However,
it is precisely this fact that, at the end, leads to the breakdown of
the existing scalar reduction methods. Namely, for some kinematics (e.g.
collinear on-shell external lines) the system (\ref{f27}) has no solution for
$C\neq 0$. However, if the possibility $C=0$ is allowed, the system (\ref{f27})
will have a solution regardless of kinematics. This makes it possible to obtain
additional reduction relations and formulate methods applicable to arbitrary
number of external lines and to arbitrary kinematics.  

If (\ref{f25}) is taken into account, 
and after using the relation \cite{davy}
\begin{equation}
-\sum\nolimits_{j=1}^{N} \nu_j I_0^N(D;\{ \nu_k
+\delta_{k j}\}) = (4 \pi \mu ^2)^{-1} I^N_0(D-2;\{ \nu_k\}),
\label{f24}
\end{equation}
which can be easily proved from the representation (\ref{f13}),
the relation (\ref{f20}) reduces to
\begin{eqnarray}
C\,I_0^N(D-2;\{\nu_k\})
&=&
\sum_{i=1}^{N}z_i I_0^N(D-2;\{\nu_k-\delta_{k i}\})
\nn \\ & &
+(4\pi \mu ^2)
(  D-1-\sum_{j=1}^{N}\nu_j ) z_0
I_0^N(D;\{\nu_k\}), \label{f31} 
\end{eqnarray}
where $z_i$ are given by the solution of the
system (\ref{f25}), and the infinitesimal part proportional to $\e$ has been
omitted.
This is a generalized form of the recursion relation
which connects the scalar
integrals in a different number of dimensions
\cite{dixon,bern,tarasov,tarasov1,binoth}.
The use of the relation (\ref{f31}) in practical calculations
depends on the form of the solution of the system of 
equations (\ref{f25}).
For general considerations, it is advantageous to write
the system (\ref{f25})
in the following way:
\begin{equation}
\left( \begin{array}{ccccc}
0 & 1 & 1 & \cdots & 1 \\
1 & 0 & r_{12} & \cdots & r_{1N} \\
1 & r_{12} & 0 & \cdots & r_{2N} \\
\vdots & \vdots & \vdots & \ddots & \vdots \\
1 & r_{1N} & r_{2N} & \cdots & 0
\end{array} \right) 
\left( \begin{array}{c}
-C \\ z_1 \\ z_2 \\ \vdots \\ z_N
\end{array} \right)
=
\left( \begin{array}{c}
z_0 \\ 0 \\ 0 \\ \vdots \\ 0
\end{array} \right) . \label{f32}
\end{equation}
In writing (\ref{f32}), we have taken into account the fact that 
$z_0=\sum_{i=1}^N z_i$. In this
way, the only free parameter is
$z_0$ and by choosing it in a convenient way, one
can always find the solution
of the above system and, consequently, be able to use
the recursion relations 
(\ref{f31}).

In the literature, for example in Ref. \cite{tarasov, tarasov1, binoth}, 
the recursion relations are obtained
by inserting the general solution of the system (\ref{f25}), i.e. 
the system (\ref{f32}), 
into the relation (\ref{f31}).
The recursion relations thus obtained are of limited practical use if the
matrices of the mentioned systems are very singular. This happens when 
there are either
two or more collinear external lines or, in general, for $N>6$.
When this is the case, the analysis of the general coefficient of the recursion
becomes very complicated and in many cases unmanageable.
There are cases when all coefficients vanish.
As stated in \cite{tarasov1}, for $N\ge 7$, owing to the drastic reduction of the recurrence
relations these cases need a separate investigation. In addition, 
the above-mentioned problems with $C \neq 0$ appear.
To avoid these problems, a different approach to recursion relations can be
taken. It is based on the fact, that finding any solution of the systems of
equations mentioned above, makes it possible to perform the reduction.
Being forced to use computers, it is very convenient and important that the
reduction procedure be organized in a such a way that the recursion relations
are classified and used depending on the form of the solutions of the above
systems. If this is done, the increased singularity of the kinematic
determinants turns out to be working in our favour by making it easy to find a
solution of the systems of linear equations relevant to the reduction.

In the following we frequently refer to two determinants,
for which we introduce the notations: for the determinant of
the system (\ref{f27}) we introduce
det$(R_N)$, while for the determinant of the system (\ref{f32}) we
 use det$(S_N)$.
Depending on whether the kinematic determinants
det$(R_N)$ and det$(S_N)$ are equal to zero or not,
we distinguish four different types of recursion
relations following from (\ref{f31}).
Before proceeding to consider various cases, note that in the
case when det$(S)\neq 0$, it holds
\begin{equation}
C=-z_0\frac{{\rm det}(R_N)}{{\rm det}(S_N)}. \label{f33}
\end{equation}

It should be mentioned that for some of the recursion relations presented 
below one can find similar expressions in the literature. For reasons of
clearness, connections of the relations given below with those existing in the
literature are commented upon after the analysis of all possible cases has been
considered.  

Let us now discuss all possible cases
separately.

\subsection{ Case I:~~
${\rm det}(S_N)\neq 0$, ${\rm det}(R_N)\neq 0$}  

The most convenient choice in this case is $z_0=1$.
It follows from (\ref{f33}) that
$C\neq 0$, so that the recursion relation 
(\ref{f31}) can be written in the following form:
\begin{eqnarray}
I_0^N(D;\{\nu_k\}) &=& \frac{1}{4 \pi \mu ^2
 (D-1-\sum\nolimits_{j=1}^{N}\nu_j )}
 \bigg[ C\,I_0^N(D-2;\{\nu_k\})\nn
 \\ & & 
-\sum\nolimits_{i=1}^{N}z_i
 I_0^N(D-2;\{\nu_k-\delta_{k i}\}) \bigg]
.\label{f34}
\end{eqnarray}
As it is seen, this recursion relation connects the 
scalar integral in $D$ dimensions with the scalar integrals in
$D-2$ dimensions and can be used to reduce the dimensionality
of the scalar integral.

Since det$(R_N)\neq 0$, some more recursion relations can be
directly derived from
(\ref{f20}).
By directly choosing the constants $z_i$ in (\ref{f20})
in such a way that
$z_i=\delta_{i k}$, for $k=1,\cdots, N$, we arrive at
a system of $N$ equations which is always valid:
\begin{eqnarray}
\sum_{j=1}^{N}
(r_k-r_j)^2 \nu_j I_0^N(D;\{\nu_i+\delta_{i j}\})
&=&\sum_{j=1}^{N} \nu_j
 I_0^N(D;\{\nu_i+\delta_{i j}-\delta_{i k}\})
 \nn \\ & & \hspace{-3cm}
-(D-\sum_{j=1}^{N}\nu_j)
 I_0^N(D;\{\nu_i\}), \qquad k=1,\cdots,N.\label{f35}
\end{eqnarray}
In the system (\ref{f35}) we have again disregarded the
non-essential infinitesimal term
proportional to $\e$. 
The matrix of the system (\ref{f35}) is the same
as the matrix of the system 
(\ref{f25}), whose determinant is different from zero,
so that
the system (\ref{f35}) can be solved with respect to 
$I_0^N(D;\{\nu_i+\delta_{i j}\})$, $j=1,\cdots,N$.
The solutions represent the recursion relations
which can be used to reduce the powers of the propagators 
in the scalar integrals.
Making use of these relations and the relation (\ref{f34}),
each scalar integral $I_0^N(D;\{\nu_i\})$ belonging to
the type for which 
det$(S_N) \neq 0$, 
det$(R_N)\neq 0$ can be represented as a linear
combination of integrals
$I_0^N(D';\{ 1\})$ and integrals with the number of 
propagators which is less than $N$.
For the dimension $D'$, one usually chooses $4+2 \eir$, 
where $\eir$ is the infinitesimal
parameter regulating the divergences.
Even in the case when one starts with $D < D'$, one can 
make use of the recursion (\ref{f34}) to change from the
dimension $D$ to the dimension $D'$.

In addition to the two sets of recursion relations 
presented above, by combining them one can obtain an additional and 
very useful set of recursion relations.
This set at the same time reduces $D$ and $\nu_i$ in all terms.
By adding and subtracting the expression
${\delta}_{jk} I_0^N(D;\{\nu_i+\delta_{ij}-\delta_{ik})$ in the 
first term on the right-hand
side of the system (\ref{f35})
and
makeing use of the relation (\ref{f24}), one finds
\begin{eqnarray}
\sum_{j=1}^{N}
(r_k-r_j)^2 \nu_j I_0^N(D;\{\nu_i+\delta_{i j}\})
&=&-(4\pi{\mu}^2)^{-1}
I_0^N(D-2;\{\nu_i-\delta_{i k}\})
 \nn \\ & & \hspace{-3cm}
-(D-1-\sum_{j=1}^{N}\nu_j)
 I_0^N(D;\{\nu_i\}), \qquad k=1,\cdots,N.\label{f37}
\end{eqnarray}
The solution of this system of equations can in principle
be used for reducing the dimension of the integral and the 
propagator powers. However, a much more useful set of the
recursion relations is obtained by combining
(\ref{f37}) and (\ref{f34}).
Expressing the second term on the right$-$hand side of
(\ref{f37}) with the help of (\ref{f34}), leads to
\begin{eqnarray}
\sum_{j=1}^{N}
(r_k-r_j)^2 \nu_j I_0^N(D;\{\nu_i+\delta_{i j}\})
&=&(4\pi{\mu}^2)^{-1}
\bigg [\sum_{j=1}^{N}(z_j-{\delta}_{jk}) 
 I_0^N(D-2;\{\nu_i-\delta_{i j}\})
 \nn \\ & & \hspace{-3cm}
-CI_0^N(D-2;\{\nu_i\})\bigg ],~~~ 
\qquad k=1,\cdots,N,\label{f38}
\end{eqnarray}
where $z_i$ and $C$ represent solutions of the
system (\ref{f32}) for $z_0=1$.
Solutions of the system (\ref{f38}) represent the recursion relations
which, at the same time, reduce (make smaller) the dimension
and the powers of the propagators in all terms (which is
very important).
As such, they are especialy convenient for making
a rapid reduction of the scalar integrals which appear
in the tensor decomposition of high-rank tensor integrals.

\subsection{ Case II:~~
${\rm det}(S_N)\neq 0$, ${\rm det}(R_N)=0$}

The most convenient choice in this case is
$z_0=1$. Unlike in the preceding case, it follows
from (\ref{f33}) that
$C=0$, so that the recursion relation
(\ref{f31}) can be written as
\begin{equation}
I_0^N(D;\{\nu_k\}) = \frac{1}{4 \pi \mu ^2
 (  D-1-\sum\nolimits_{j=1}^{N}\nu_j )}\left[ -
 \sum\nolimits_{i=1}^{N}z_i I_0^N(D-2;\{\nu_k-
\delta_{k i}\}) \right]
.\label{f39}
\end{equation}
It follows from (\ref{f39}) that it is possible 
to represent each integral of this type as a linear
combination of scalar integrals with the number of 
propagators being less than $N$.

\subsection{ Case III:~~
${\rm det}(S_N)=0$, ${\rm det}(R_N)\neq 0$}

This possibility arises only if the first row of the
matrix of the system (\ref{f32}) is a linear combination
of the remaining rows. 
Then, the system (\ref{f32}) has a solution 
only for the choice $z_0=0$.
With this choice, the remaining system of equations 
reduces to the system
(\ref{f25}), where the constant $C$ can be chosen 
at will.
After the parameter $C$ is chosen, the constants  $z_i$ 
are uniquely determined.
Thus the recursion relation (\ref{f31}) with the choice $C=1$ leads to 
\begin{equation}
I_0^N(D;\{\nu_k\}) = 
\sum\nolimits_{i=1}^{N}z_i I_0^N(D;\{\nu_k-\delta_{k i}\})
.\label{f40}
\end{equation}
Consequently, as in the preceding case, the scalar integrals 
of the type considered
can be represented as a linear combination of scalar integrals 
with a smaller number of propagators.

\subsection{ Case IV:~~
${\rm det}(S_N)=0$, ${\rm det}(R_N)=0$}

Unlike in the preceding cases, in this case
two different recursion relations arise.
To derive them, we proceed by subtracting the last,
$(N+1)$$-$th, equation of the system (\ref{f32}) from the
second, third,... and $N$$-$th equation, respectively.
As a result, we arrive at the following 
system of equations:
\begin{eqnarray}
& &\left( \begin{array}{ccccc}
0 & 1 & 1 & \cdots & 1 \\
0 & -r_{1N} & r_{12}-r_{2N} & \cdots & r_{1N} \\
0 & r_{12}-r_{1N} & -r_{2N} & \cdots & r_{2N} \\
\vdots & \vdots & \vdots & \ddots & \vdots \\
0 & r_{1,N-1}-r_{1N} & r_{2,N-1}-r_{2N} & \cdots & 
r_{N-1,N} \\
1 & r_{1N} & r_{2N} & \cdots & 0
\end{array} \right)  
\left( \begin{array}{c}
-C \\ z_1 \\ z_2 \\ \vdots \\ z_{N-1} \\ z_N
\end{array} \right)
=
\left( \begin{array}{c}
z_0 \\ 0 \\ 0 \\ \vdots \\ 0 \\ 0
\end{array} \right) .~~~~~~ \label{f41}
\end{eqnarray}
As it is seen, the first $N$ equations of the above system
form a system of equations in which the constant $C$ does
not appear, and which can be used to determine the constants 
$z_i$, $i=1,\cdots N$.
The fact that det$(S_N)=0$ implies that the determinant
of this system vanishes. Therefore, for the system
in question to be consistent (for the solution to exist),
the choice $z_0=0$ has to be made. Consequently, the solution
of the system, $z_i\,\,(i=1,2,...N)$, will contain at least
one free parameter. Inserting this solution into the last,
$(N+1)$$-$th, equation of the system (\ref{f41}), we obtain 
\begin{equation}
\sum\nolimits_{i=1}^{N} r_{iN} z_i=C~. \label{f42}
\end{equation}
Now, by arbitrarily choosing the parameter $C$, one of 
the free parameters on the left-hand side can be fixed.

Sometimes, (for instance, when there are 
collinear external lines) the left-hand side
of Eq. (\ref{f42}) vanishes explicitly, although the solution
for $z_i$ contains free parameters.
In this case the choice $C=0$ has to be made.

Therefore, corresponding to the case when
det$(S_N)$=det$(R_N)=0$, 
one of the following two recursion relations holds:
\begin{equation}
I_0^N(D;\{\nu_k\}) = 
\sum\nolimits_{i=1}^{N}z_i I_0^N(D;\{\nu_k-\delta_{k i}\})
,\label{f43}
\end{equation}
obtained from (\ref{f31}) by setting $z_0=0$ and $C=1$, or
\begin{equation}
0 = 
\sum\nolimits_{i=1}^{N}z_i I_0^N(D;\{\nu_k-\delta_{k i}\})
,\label{f44}
\end{equation}
obtained from (\ref{f31}) by setting $z_0=0$ and $C=0$.
 
In the case (\ref{f43}), it is clear that the integral
with $N$ external lines can be represented in terms of the 
integrals with $N-1$ external lines.
What happens, however, in the case (\ref{f44})?
With no loss of generality, we can take that $z_1\neq 0$.
The relation (\ref{f44}) can then be written in the form
\begin{equation}
z_1 I_0^N(D;\{\nu_k\}) = 
-\sum\nolimits_{i=2}^{N}z_i I_0^N(D;\{\nu_k+\delta_{k 1}
-\delta_{k i}\})
.\label{f45}
\end{equation}
We can see that, in this case too, the integral 
with $N$ external lines can be represented in terms of the 
integrals with $N-1$ external lines.
In this reduction, 
$\sum\nolimits_{i=1}^{N} \nu_i$ remains conserved.

Based on the above considerations, it is clear that
in all the above cases with the exception of that when
det$(S_N) \neq 0$, det$(R_N) \neq 0$, the integrals with
$N$ external lines can be represented in terms of the 
integrals with smaller number of external lines.
Consequently, then, there exists a fundemantal set of
integrals in terms of which all integrals can be
represented as a linear combination.


Before moving on to determine a fundamental set of integrals, let us briefly
comment on the recursion relations for scalar integrals that can be found in the
literature.
As we see below, det$(S_N)$ is proportional to the Gram determinant. All
recursion relations for which the Gram determinant does not vanish are well
known. Thus the relation of the type  (\ref{f34}) 
can be found in Refs. \cite{dixon,bern,tarasov,tarasov1,binoth},
while the solutions of the systems (\ref{f35}) and (\ref{f38}) 
correspond to the recursion
relations (28) and (30), respectively, given in Ref. \cite{tarasov1}. 
Even though Case
II also belongs to the class of cases for which the Gram determinant is
different from zero, the system (\ref{f27}) 
has no solution for
$C\neq 0$.
This is a reason why the problem with using recursion relations appear in all
approaches in which it is required that $C\neq 0$. This can be seen from the
discussion in \cite{binoth} (the method is based on the choice $C=1$) where 
the authors
state that the reduction cannot be done for $N=3$ with on-shell external lines,
and for $N=4$ when one of the Mandelstam variables $s$ or $t$ vanishes.
Such cases, however, are unavoidable when obtaining 
leading twist NLO PQCD predictions for
exclusive processes at large momentum transfer.
On the other hand, in the approach 
of Ref. \cite{tarasov1},
where all coefficients of the recursion
are given in terms of det$(S_N)$ and the minors of the matrix $S_N$, 
the relation of the type (\ref{f39}) can be obtained 
(Eq. (35) in Ref. \cite{tarasov1}).
Cases III and IV, for which the Gram determinant vanishes are of special
interest. One of the most discussed cases in the literature, belonging to Case
III, is $N=6$. The recursion relations of the type (\ref{f40}) can be found in Ref.
\cite{bern,tarasov,tarasov1,binoth}. 
As for Case IV, it is especially interesting owing to the fact
that it includes all cases for $N\ge 7$. In this case, the systems (\ref{f27})
 and (\ref{f32})
have no unique solution.
Case IV causes a lot of trouble for approaches in which the recursion
coefficients are given in terms of det$(S_N)$ and the minors of the matrix 
$S_N$, for example in Ref. \cite{tarasov1}. 
The problem consists in the fact that all
determinants vanish, making it impossible to formulate the recursion
relation, so these cases need a separate investigation. On the other hand, the
method of Ref. \cite{binoth}, based on using pseudo-inverse matrices, 
can be used to
construct the most general solution of the system (\ref{f27}) 
for the case of the vanishing
Gram determinant. Even though the authors of Ref. \cite{binoth} 
claim that using their
approach one can always perform the reduction of the $N$-point function ($N\ge
6$), that does not seem to be the case.
Namely, the method in \cite{binoth} is based on the choice 
$C=1$, and as it has been shown
above, in some cases belonging to Case IV the system (\ref{f25}) 
has no solution
for $C\neq 0$, implying that the reduction cannot be performed.
The impossibility of performing the reduction manifests itself such that $v\cdot
K\cdot v =0$ (see (15) and (19) in \cite{binoth}), 
a consequence of which is that the
recursion coefficients become divergent.
The situation of this kind arises regularly when dealing with integrals containing
collinear external lines, i.e. for exceptional kinematics. The method of Ref.
\cite{binoth} has been obtained for non-exceptional kinematics.

In view of what has been said above, most of the problems with existing
reduction methods appear when dealing with the integrals with a large number of
external lines.
In all considerations in the literature that happens for $N>6$. It is very
important to point out that, this is valid for the case when external momenta span
the 4-dimensional Minkowski space. If the dimensionality of space span by
external momenta is smaller, the problems start appearing for smaller $N$.
Even though one can find the statements that such cases are at the moment of
minor physical interest, we disagree.
Namely, as stated earlier, the analysis of exclusive processes in PQCD, even for
simple processes, requires evaluation of the diagrams with $N\ge 6$.
Since these diagrams contain collinear external lines, the kinematics is limited
to ($d<4$)-dimensional subspace.
Thus, for example, for nucleon Compton scattering the integrals with $N=8$
external lines contribute and the kinematics is limited to the 3-dimensional
subspace. A consequence of this is that the problem with using existing
reduction methods will start appearing at the level of
one-loop $N=5$ diagrams.

The reduction method presented in this paper is formulated with an eye on
exclusive processes in PQCD. 
The main point of the method is that the reduction is defined in terms of the
solution of the linear systems given by (\ref{f27})
 and (\ref{f32}).
A consequence of this is that the method is quite general, very flexible,
practical and easily transfered to the computer program.
To perform reduction, 
one only needs to find solution of the above systems which can always be done.
A very pleasing feature of this reduction is that the increased 
singularity of
the kinematic determinants  facilitates reduction, since finding 
a solution of the
relevant linear systems becomes easy.

\section{On the fundamental set of integrals}

We now turn to determine the fundamental set of integrals.
To this end, let us first evaluate the determinant of
the system (\ref{f32}), det$(S_N)$, and determine the conditions
under which this kinematic determinant vanishes.

By subtracting the last column from the second, third, ...
and $N-$th column, respectively, and then the last
row from the second, third, ... and $N-$th row,
respectively, we find that the determinant det$(S_N)$
is given by the following expression:
\begin{equation}
{\rm det}(S_N)=-{\rm det}\left[-2(r_i-r_N)(r_j-r_N)\right],
\qquad i,j=1,\ldots,
N-1. \label{f46}
\end{equation} 
As it can be seen, det$(S_N)$ is proportional to the Gram determinant.
Denote by $n$ the dimension of the vector space
spanned by the vectors $r_i-r_N$,~ ($i=1,\ldots, N-1$).
Owing to the linear dependence of these vectors,
the determinant vanishes when $N>n+1$.
As in practice, we deal with the 4$-$dimensional 
Minkowski space, the maximum value for $n$ equals 4.
An immediate consequence of this is that all integrals 
with $N>5$ can be reduced to the integrals with $N\leq 5$.

In view of what has been said above, all one-loop
integrals are expressible in terms of the integrals
$I_0^3(4+2\eir;\{1\})$,
$I_0^4(4+2\eir;\{1\})$,
$I_0^5(4+2\eir;\{1\})$,
belonging to Case I, and the general 
two$-$point integrals
$I_0^2(D';{\nu}_1,{\nu}_2)$,
which are simple enough to be evaluated analytically.
 
Next, by substituting 
$D=6+2 \eir$, $N=5$ and ${\nu}_i=1$
into the recursion relation (\ref{f31}), one finds
\begin{equation}
C\,I_0^5(4+2\eir;\{1\})
=
\sum_{i=1}^{5}z_i 
I_0^5(4+2\eir;\{\delta_{k k}-\delta_{k i}\})
+(4\pi \mu^2)
(2{\eir})~z_0~I_0^5(6+2\eir;\{1\}). \label{f47}
\end{equation}
Owing to the fact that the integral
$I_0^5(6+2\eir;\{1\})$ is IR finite\cite{bern},
the relation (\ref{f47}) implies that the
$N=5$ scalar integral, $I_0^5(4+2\eir;\{1\})$,
can be expressed as a linear combination of
the $N=4$ scalar integrals, $I_0^4(4+2\eir;\{1\})$,
plus a term linear in $\eir$.
In massless scalar theories, the term 
linear in $\eir$ can simply be omitted,
with a consequence that the $N=5$ integrals
can be reduced to the $N=4$ integrals.
On the other hand, when calculating in  
renormalizable gauge theories (like QCD),
the situation is not so simple,
owing to the fact that the rank 
$P(\le N)$ tensor integrals are required.

In the process of the tensor decomposition
and then reduction of scalar integrals all way down
to the fundamental set of integrals, there appears
a term of the form
$(1/{\eir}) I_0^5(4+2\eir;\{1\})$, 
which implies that one would need to know an
analytical expression for the integral
$I_0^5(4+2\eir;\{1\})$,
to order $\eir$.
Going back to the expression (\ref{f47}), we notice that
all such terms can be written as a linear combination
of the box (4$-$point) integrals in $4+2 \eir$ dimensions
and 5$-$point integrals in $6+2\eir$ dimensions.
Therefore, at this point, the problem has been reduced
to calculating the integral
$I_0^5(6+2\eir;\{1\})$,
which is IR finite, and need to be calculated to
order ${\cal O}({\eir}^{0})$.
It is an empirical fact \cite{bern,tarasov,tarasov1,binoth,tarnew} 
that in final expressions
for physical quantities all terms containing the integral
$I_0^5(6+2\eir;\{1\})$
always combine so that this integral ends up being
multiplied by the coefficients
$\cal O(\eir)$, and as such, can be omitted in
one$-$loop calculations.
A few theoretical proofs of this fact can be found in literature
\cite{bern,tarasov1,binoth}, but, to the best of our knowledge, 
the proof for the case of exceptional kinematics is still missing.
That being the case, in concrete calculations,
(to be sure and to have all the steps of the calculation under control
),
it is absolutely necessary to keep track of all the
terms containing
the integral $I_0^5(6+2\eir;\{1\})$, add them up and check whether the factor
multiplying it is of order ${\cal O}(\eir)$.
Even though the experience gained in numerous calculations shows that this is
so, a situation in which the integral $I_0^5(6+2\eir;\{1\})$ would appear in the
final result for a physical quantity accompained by a factor ${\cal O}(1)$
would not, from
practical point of view, present any problem. 
Namely, being IR finite, although extremely complicated to be evaluated
analyticaly, the integral $I_0^5(6+2\eir;\{1\})$ can always, 
if necessary, be evaluated numerically.

Based on the above considerations we may conclude
that all one$-$loop integrals occurring when evaluating
physical processes in massless field theories can be 
expressed in terms of the integrals
\[
I_0^2(D';{\nu}_1,{\nu}_2),~~~
I_0^3(4+2\eir;1,1,1),~~~
I_0^4(4+2\eir;1,1,1,1).
\]
These integrals, therefore, constitute a minimal set
of fundamental integrals.

In view of the above discussion, we conclude that the
set of fundamental integrals  is comprised
of integrals with two, three and four external lines.
Integrals with two external lines can be calculated
analytically in arbitrary number of dimensions
and with arbitrary powers of the propagators.
They do not constitute a problem.
As far as the integrals with three and four external
lines are concerned,
depending on how many kinematic variables vanish,
we distinguish several
different cases.
We now show that in the case $N=3$ we have only one 
fundamental integral, while in the case corresponding 
to $N=4$ there are six integrals.
For this purpose, we make use of the vanishing of the kinematic
determinants det$(R_N)$ and det$(S_N)$.
\subsection{The general scalar integral for $N=2$}
According to (\ref{f5}), the general massless scalar two$-$point
integral in $D$ space$-$time dimensions is of the form
\begin{equation}
I_0^2(D;\nu_1,\nu_2)\equiv
(\mu ^2)^{2-D/2}\int
\frac{{\rm d}^D l}{(2\pi)^D}
\frac{1}
{A_1^{{\nu}_1}
A_2^{{\nu}_2}} \cdot \label{f48}
\end{equation}
The closed form expression for the above integral, valid
for arbitrary $D=n+2\eir$, and arbitrary propagator powers 
${\nu}_1$ and ${\nu}_2$, is given by 
\begin{eqnarray}
I^2_0(n+2\eir;\nu_1, \nu_2)&=&
(4\pi \mu^2)^{2-n/2}(-1)^{\nu_1+\nu_2}\left 
(-p^2-\e \right)^{n/2-\nu_1-\nu_2}
\nn \\ 
&\times &
\frac
{\Gamma\left (\nu_1+\nu_2-n/2-\eir\right)}
{\Gamma(-\eir)}
\frac
{\Gamma\left(n/2-\nu_1+\eir\right)}
{\Gamma(1+\eir)}
\frac
{\Gamma\left(n/2-\nu_2+\eir\right)}
{\Gamma(1+\eir)}
\nn \\ 
&\times &
\frac
{1}
{\Gamma(\nu_1)\Gamma(\nu_2)}
\frac
{\Gamma(2+2\eir)}
{\Gamma(n-\nu_1-\nu_2+2\eir)}
I_0^2(4+2\eir,1,1)
\, , \label{f49}
\end{eqnarray}
where
\begin{eqnarray}
I^2_0(4+2\eir;1,1)&=&
\frac
{{\rm i}}{(4\pi)^2}
\left (
-
\frac
{p^2+\e}
{4\pi \mu ^2}
\right )^{\eir}
\frac
{\Gamma \left(-\eir \right) 
\Gamma^2\left(1+\eir \right)}
{\Gamma\left( 2+2\eir \right)}
\, . \label{f50}
\end{eqnarray}
It is easily seen that in the formalism of the dimensional
regularization the above integral vanishes for $p^2=0$.
\subsection{The scalar integrals for $N=3$}
The massless scalar one$-$loop triangle integral in
$D=4+2\eir$ dimensions is given by
\begin{equation}
I_0^{3}(4+2\eir,\{1\})=({\mu}^2)^{-\eir}
\int \frac{{\rm d}^{4+2\eir}l}{(2\pi)^{4+2\eir}}\frac{1}
{A_1 A_2 A_3 }~.\label{f51}
\end{equation}
Making use of the representation (\ref{f13}), 
and introducing the external 
masses $p_i^2=m_i^2$ 
$(i=1,2,3)$,
the integral (\ref{f51}) can be written in the form
\begin{eqnarray}
I_0^3(4+2\eir,\{1\})&=& 
\frac{{-\rm i}}{(4\pi)^2}
\frac{\Gamma (1-\eir)}
{(4\pi {\mu}^2)^{\eir}}
\int_0^1{\rm d}x_1 {\rm d}x_2 
{\rm d}x_3 
\;\delta(x_1+x_2+x_3-1)
\nonumber \\
& &\times 
\left(-x_1x_2\;m_2^2-x_2x_3\;m_3^2-x_3x_1\;m_1^2
-{\rm i}\epsilon\right)^{\eir -1}. 
\label{f52}
\end{eqnarray}
It is evident that the above integral is invariant
under permutations of external massess $m_i^2$.
Depending on the number of the external massless lines,
and using the above mentioned symmetry,
there are three relevant special cases of the above integral.
We denote them by
\begin{eqnarray}
I_3^{1m} &\equiv & I_0^3(4+2\eir,\{1\};0,0,m_3^2),
\label{f53} \\
I_3^{2m} &\equiv & I_0^3(4+2\eir,\{1\};0,m_2^2,m_3^2),
\label{f54} \\
I_3^{3m} &\equiv & I_0^3(4,\{1\};m_1^2,m_2^2,m_3^2),
\label{f55} 
\end{eqnarray}
The integrals $I_3^{1m}$ and $I_3^{2m}$ are IR divergent 
and need to be evaluated with $\eir > 0$,
while the integral $I_3^{3m}$ is finite and
can be calculated with $\eir = 0$.

Now, it is easily found that the determinants of
the systems of equations (\ref{f27}) and (\ref{f32}) are, for $N=3$,
given by 
\begin{eqnarray}
{\rm det}(R_3)&=&2 m_1^2m_2^2m_3^2,\label{f56}\\
{\rm det}(S_3)&=&(m_1^2)^2+(m_2^2)^2+(m_3^2)^2
-2 m_1^2m_2^2-2 m_1^2m_3^2-2m_2^2m_3^2, \label{f57} 
\end{eqnarray}
As is seen from (\ref{f56}), if at least one of the external
lines is on mass-shell, the determinant det$(R_3)$
vanishes. Consequently, using the recursion relations
(Case II or IV)
the integrals $I_3^{1m}$ and $I_3^{2m}$ can be reduced to
the integrals with two external lines.
Therefore, we conclude that among the scalar integrals
with three external lines the integral $I_3^{3m}$ is the
only fundamental one.
 
The result for this integral is well known \cite{mi2,davy1,three}.
In \cite{mi2} it is expressed in terms of the
dimensionless quantities of the form
\begin{equation}
x_{1,2}=
\frac{1}{2}
\left [
1-\frac{m_1^2}{m_2^2}+\frac{m_3^2}{m_2^2}
\pm
\sqrt
{\left (1-\frac{m_1^2}{m_2^2}-\frac{m_3^2}{m_2^2}\right)^2-
4 \frac{m_1^2}{m_2^2}\frac{m_3^2}{m_2^2}}
\right] \label{f58}
\end{equation}
and, being proportional to $1/(x_1-x_2)$, appears to have
a pole at $x_1=x_2$. It appears that the final expression \cite{mi2}
is not well defined when $x_1=x_2$.

On the basis of Eqs.(\ref{f57}) and (\ref{f58}), one finds that
\[
x_1-x_2=\frac{1}{m_2^2}
\sqrt {{\rm \det}(S_3)}.
\]
This equation implies that when $x_1-x_2=0$,
instead of examining the limit of the general expression
in \cite{mi2}, one can utilize the reduction relations (\ref{f40})
(corresponding to ${\rm det}(R_3)\ne 0$ and
${\rm det}(S_3)=0$) to reduce the IR finite integral
with three external lines, $I_3^{3m}$, to the integrals
with two external lines.
\subsection{The scalar integrals for $N=4$}
The massless scalar one$-$loop box integral in
$D=4+2\eir$
space$-$time dimensions is given by
\begin{equation}
I_0^{4}(4+2\eir,\{1\})=({\mu}^2)^{-\eir}
\int \frac{{\rm d}^{4+2\eir}l}{(2\pi)^{4+2\eir}}\frac{1}
{A_1 A_2 A_3 A_4}~,\label{f59}.
\end{equation}
Making use of (\ref{f13}),
introducing the external "masses"
$p_i^2=m_1^2$ $(i=1,2,3,4)$, 
and the Mandelstam variables
$s=(p_1+p_2)^2$ and $t=(p_2+p_3)^2$, the integral
(\ref{f59}) becomes
\begin{eqnarray}
\lefteqn{I_0^4(4+2\eir,\{1\})= 
\frac{{\rm i}}{(4\pi)^2}
\frac{\Gamma (2-\eir)}
{(4\pi {\mu}^2)^{\eir}}
\int_0^1{\rm d}x_1 {\rm d}x_2 
{\rm d}x_3 {\rm d}x_4 
\;\delta(x_1+x_2+x_3+x_4-1)}
\nonumber \\
& &\times \left(-x_1x_3\;t-x_2x_4\;s-x_1x_2\;m^2_2\right . 
\left .-x_2x_3\;m^2_3
-x_3x_4\;m^2_4-x_1x_4\;m^2_1-{\rm i}\epsilon\right)^{\eir -2}. 
\label{f60}
\end{eqnarray}
Introducing the following set of ordered pairs
\begin{equation}
{(s,t),\,(m_1^2,m_3^2),\,(m_2^2,m_4^2)},
\label{f61}
\end{equation}
one can easily see that the integral (\ref{f60})
is invariant under the permutations of ordered pairs,
as well as under the simultaneous exchange of places of 
elements in any two pairs.

The determinants of the coefficient matrices of
the systems of equations 
(\ref{f27}) and (\ref{f32}), corresponding to the above integral, are
\begin{eqnarray}
{\rm det}(R_4)
&=& {} 
s^2t^2+(m_1^2m_3^2)^2+(m_2^2m_4^2)^2
\nonumber \\
& & {}
-2stm_1^2m_3^2-2stm_2^2m_4^2
-2m_1^2m_2^2m_3^2m_4^2\,. \label{f62}\\
& & {} 
\nonumber \\
{\rm det}(S_4)
&=& {} 
2\,[\,st\,(m_1^2+m_2^2+m_3^2+m_4^2-s-t)
\nonumber \\
& & {} 
+m_2^2m_4^2(s+t+m_1^2+m_3^2-m_2^2-m_4^2)
\nonumber \\
& & {} 
+m_1^2m_3^2(s+t-m_1^2-m_3^2+m_2^2+m_4^2)
\nonumber \\
& & {} 
-s(m_1^2m_2^2+m_3^2m_4^2)-t(m_1^2m_4^2+m_2^2m_3^2)\,]~.\label{f63}
\end{eqnarray}

By looking at the expression for ${\rm det}(R_4)$ given in
(\ref{f62}) it follows that all box integrals $I_0^4$ that are
characterized by the fact that in each of the ordered pairs
in (\ref{f61}) at least one kinematic variable vanishes, are reducible.
Therefore, for a box integral to be irreducible, it is necessary
that both kinematic variables in at least one of the
ordered pairs should be different from zero.
Owing to the symmetries valid for the box integrals
it is always possible to choose that pair to be $(s,t)$.

Taking into account symmetries,
and the number of  
external massless lines, 
there are six potentially irreducible
special cases of the integral (\ref{f60}).
Adopting the notation of Ref. \cite{bern}, we denote them by
\begin{eqnarray}
I_4^{4m} &\equiv & I_0^4(4,\{1\};s,t,m_1^2,m_2^2,m_3^2,m_4^2), 
\label{f64} \\
I_4^{3m} &\equiv & I_0^4(4+2\eir,\{1\};s,t,0,m_2^2,m_3^2,m_4^2),
 \label{f65} \\
I_4^{2mh} &\equiv & I_0^4(4+2\eir,\{1\};s,t,0,0,m_3^2,m_4^2),
 \label{f66} \\
I_4^{2me} &\equiv & I_0^4(4+2\eir,\{1\};s,t,0,m_2^2,0,m_4^2),
 \label{f67} \\
I_4^{1m} &\equiv & I_0^4(4+2\eir,\{1\};s,t,0,0,0,m_4^2),
 \label{f68} \\
I_4^{0m} &\equiv & I_0^4(4+2\eir,\{1\};s,t,0,0,0,0,),
 \label{f69} 
\end{eqnarray}
with all kinematic variables appearing above being
different from zero. 
The results for these integrals are well known \cite{bern,mi1,mi2,box}.

The integrals (\ref{f65}-\ref{f69}) are IR divergent,
and as such need to be evaluated  with $\eir > 0$, while the
integral (\ref{f64}) is finite and can be calculated in $D=4$.
The results for these integrals, 
obtained in \cite{mi1,mi2} for arbitrary values of the relevant
kinematic variables, and presented in a simple and compact
form, have the following structure:
\begin{eqnarray}
I_4^K(s,t;{m_i^2})
&=& {} 
\frac{i}{(4{\pi})^2}\,
\frac
{\Gamma (1-\eir){\Gamma}^2(1+\eir)}
{\Gamma (1+2\eir)}\,
\frac{1}{\sqrt {{\rm det}(R_4^K)}}
\nonumber \\
& & {}
\times
\left [
\frac{G^K(s,t;\eir;{m_i^2})}{{\eir}^2}
+H^K(s,t;{m_i^2})
\right ]
+{\cal O}(\eir),
\nonumber \\
& & {} 
K\in \{0m,1m,2me,2mh,3m,4m\}\,. \label{f70}
\end{eqnarray}
The IR divergences (both soft and collinear) of the 
integrals are contained in the first term within the
square brackets, while the second term is finite.
The function
$G^K(s,t;\eir;{m_i^2})$
is represented by a sum of powerlike terms, it depends
on $\eir$ and is finite in the 
$\eir \rightarrow 0$ limit.
As for the function
$H^K(s,t;{m_i^2})$, it is given in terms of dilogarithm
functions and constants.
In the above, ${\rm det}(R_4^K)$
is the determinant corresponding to the integral
$I_4^K$ given in (\ref{f64}-\ref{f69}).

For the purpose of numerical integration, it is very useful
to have the exact limit of the integral $I_4^K$ when
${\rm det}(R_4^K)\rightarrow 0$. 
This limit can be determined in an elegant 
manner by noticing that for
${\rm det}(R_4^K)= 0$
the reduction relations corresponding to Cases II and 
IV apply, making it possible
to represent the box integral $I_4^K$ as a linear 
combination of the triangle integrals.
This result can be made use of to combine box 
and triangle integrals (or pieces of these integrals)
with the aim of obtaining numerical stability of the
integrand \cite{glover}.

The integrals (\ref{f64}-\ref{f69}) are irreducible only 
if the corresponding kinematic determinant
${\rm det}(R_4^K)$ does not vanish.

With the help of the tensor decomposition
and the scalar reduction procedures, any 
dimensionally regulated one$-$loop
$N-$point Feynman integral can be represented
as a linear combination of the integrals:
\begin{eqnarray}
& & I_0^2(D';{\nu}_1,{\nu}_2),\nn \\
& & I_3^{3m}, \nn \\
& & I_4^{4m},~
I_4^{3m},~
I_4^{2mh},~
I_4^{2me},~
I_4^{1m},~
I_4^{0m},\label{f71}
\end{eqnarray}
multiplied by tensor structures made from the external
momenta and the metric tensor. 
The integrals in (\ref{f71}) constitute a fundamental set of
integrals.
An alternative and more convenient set of fundamental
integrals is obtained by noticing that all the relevant
box integrals are finite in $D=6$.
On the basis of Eq. (\ref{f31}), all IR divergent box
integrals can be expressed as linear combinations of
triangle integrals in $D=4+2\eir$ dimension and a box
integral in $D=6+2\eir$ dimension.
Next, using the same equation, all triangle integrals
can be decomposed into a finite triangle integral and 
two$-$point integrals.
In the final expression thus obtained all divergences,
IR as well as UV, are contained in the general two-point
integrals and associated coefficients.
Therefore, an alternative fundamental set of 
integrals is comprised of
\begin{eqnarray}
& & I_0^2(D';{\nu}_1,{\nu}_2),\nn \\
& & I_3^{3m}, \nn \\
& & I_4^{4m},~
J_4^{3m},~
J_4^{2mh},~
J_4^{2me},~
J_4^{1m},~
J_4^{0m},\label{f72}
\end{eqnarray}
where $J_4^K$ denotes box integrals in $D=6$ dimensions,
explicit expressions for which are given in the Appendix A.
A characteristic feature of this fundamental set of integrals, which makes it 
particularly interesting, is that the integral $I_0^2$ is the only divergent
one, while the rest of integrals are finite.

\section{Conclusion}
\label{sec:con}

In this work we have considered one$-$loop scalar and
tensor Feynman integrals with an arbitrary number of
external lines which are relevant for construction of
multi$-$parton one$-$loop amplitudes in massless 
field theories.


Main result of this paper is a scalar reduction approach
by which an arbitrary
$N-$point scalar one$-$loop integral can be
reqursively represented as a linear combination of
eight basic scalar integrals
with rational coefficients depending on the external momenta
and the dimensionality of space$-$time,
provided the external momenta are kept in four dimensions.
The problem of vanishing of the kinematic determinants,
which is a reflection of very complex singularity structure
of these integrals,
has been solved in an elegant and transparent manner.
Namely, the approach has been taken according to which
instead of solving the general system of linear equations
given in (\ref{f25}), and then finding the limit, which
sometimes doesn't exists, of the obtained
solution corresponding to a given singular kinematic
situation, we first obtain and then solve the system of
equations appropriate to the situation being considered.

Our method
has been derived without any restrictions regarding the
external momenta.
As such, it is completely general and applicable for arbitrary kinematics.
In particular, it applies to the integrals in which the
set of external momenta contains subsets comprised of two
or more collinear momenta.
This kind of integrals are encountered when performing
leading$-$twist NLO 
PQCD analysis of the hadronic exclusive processes at
large$-$momentum$-$transfer.
Trough the tensor decomposition and scalar reduction
presented, any massless one-loop Feynman integral
with generic 4-dimensional momenta can
be expressed as a linear combination of a fundamental
set of scalar integrals: six box integrals in $D=6$,
a triangle integral in $D=4$, and a general two$-$point
integral. All the divergences present in the original
integral are contained in the general two-point integral
and associated coefficients.

In conclusion, the computation of IR divergent one$-$loop
integrals for arbitrary number of external lines can be 
mastered with
the reduction formulas presented above.
The iterative structure makes it easy to
implement the formalism in algebraic computer program.
With this work all the conceptual problems concerning 
the construction of 
multi$-$parton one$-$loop amplitudes are thus solved.

%
%

\section*{Acknowledgements}
We thank T.Binoth and G.Heinrich for useful discussions and helpful 
comments.
This work was supported by the Ministry of Science and Technology
of the Republic of Croatia under Contract No. 00980102.

\section*{Appendix A}
In addition to the explicit calculation, the irreducible box 
integrals in $D=6$ dimensions can be obtained using the
existing analytical expressions for the irreducible box
integrals in $D=4+2\eir$ dimensions and the reduction
formula (\ref{f31}).
To this end, we substitute 
$D=6+2\eir$, $N=4$, ${\nu}_i=1$ and $C=1$
into the relation (\ref{f31}) and find
\begin{equation}
I_0^4(6+2\eir;\{1\}) =
\frac{1}{4\pi \mu^2(2{\eir}+1)~z_0}
\left (I_0^4(4+2\eir;\{1\})-
\sum_{i=1}^{4}z_i 
I_0^4(4+2\eir;\{\delta_{k k}-\delta_{k i}\})\right)~.\label{f73}
\end{equation}

Note that the IR divergences in $D=4+2\eir$ box integrals
are exactly cancelled by the divergences of the triangle
integrals.

The expressions for the relevant basic
massless scalar box integrals in $D=6$ space$-$time
dimensions are listed below:

\noindent
\vspace{0.5cm}
\underline {The three$-$mass scalar box integral}
\begin{eqnarray}
\lefteqn{I^{3m}_{4}(D=6; s,t;m_2^2,m_3^2,m_4^2)=
\frac{{\rm i}}{(4\pi)^{2}}\,\frac{1}{4 \pi \mu^2}\,}
\nonumber \\
& & {}\times
h^{3m}\Bigg\{ 
\frac{1}{2} \ln \left( \frac{s+{\rm i}\epsilon}
{m_3^2+{\rm i}\epsilon}\right)
\ln \left( \frac{s+{\rm i}\epsilon}
{m_4^2+{\rm i}\epsilon}\right)+
\frac{1}{2} \ln \left( \frac{t+{\rm i}\epsilon}
{m_2^2+{\rm i}\epsilon}\right)
\ln \left( \frac{t+{\rm i}\epsilon}
{m_3^2+{\rm i}\epsilon}\right)
\nonumber \\
& &
{}+\mbox{Li}_{2}\left(\, 1-\frac{m_{2}^{2}+{\rm i}\epsilon }
{t+{\rm i}\epsilon}\,
\right)+\mbox{Li}_{2}\left(\, 1-\frac{m_{4}^{2}+
{\rm i}\epsilon}
{s+{\rm i}\epsilon}\, \right)
\nonumber \\
& &
{}+\mbox{Li}_{2}\Big[\, 1-(s+{\rm i}\epsilon)\, f^{3m}\,\Big]
+\mbox{Li}_{2}\Big[\,
1-(t+{\rm i}\epsilon)\, f^{3m}\, \Big]
\nonumber \\
& &
{}-\mbox{Li}_{2}\Big[\, 1-(m_{2}^{2}+
{\rm i}\epsilon)\, f^{3m}\, \Big]
-
\mbox{Li}_{2}\Big[\,
1-(m_{4}^{2}+{\rm i}\epsilon)\, f^{3m}\, 
\Big] \nonumber \\
& &
{}-\frac{1}{2} \left( t-m_2^2-m_3^2+2 m_2^2 m_3^2
\frac{t-m_4^2}{s t-m_2^2 m_4^2}\right)
{\cal I}_3(m_2^2,m_3^2,t)\nonumber \\
& &
{}-\frac{1}{2} \left( s-m_3^2-m_4^2+2 m_3^2 m_4^2
\frac{s-m_2^2}{s t-m_2^2 m_4^2}\right){\cal I}_3(m_3^2,m_4^2,s)
\Bigg\}
.\label{f74}
\end{eqnarray}
\vspace{0.5cm}
\underline {The adjacent ("hard") two$-$mass 
scalar box integral} 
\begin{eqnarray}
\lefteqn{I_4^{2mh}(D=6; s,t;m_3^2,m_4^2) =
\frac{{\rm i}}{(4\pi)^{2}}\,\frac{1}{4 \pi \mu^2}\,
}\nonumber \\ 
& & {}\times h^{2mh}\Bigg\{ 
\frac{1}{2} \ln \left( \frac{s+{\rm i}\epsilon}
{m_3^2+{\rm i}\epsilon}\right)
\ln \left( \frac{s+{\rm i}\epsilon}
{m_4^2+{\rm i}\epsilon}\right)+
\mbox{Li}_{2}\left(\, 1-\frac{m_{4}^{2}+
{\rm i}\epsilon }{s+{\rm i}\epsilon}\,
\right)
\nonumber \\
& &
{}-\mbox{Li}_{2}\left(\, 1-\frac{m_{3}^{2}+{\rm i}\epsilon}
{t+{\rm i}\epsilon}\, \right)
+\mbox{Li}_{2}\Big[\, 1-(s+{\rm i}\epsilon)\, f^{2mh}\, \Big]
\nonumber \\
& &
{}+\mbox{Li}_{2}\Big[\,
1-(t+{\rm i}\epsilon)\, f^{2mh}\,  \Big]
-\mbox{Li}_{2}\Big[\, 1-(m_{4}^{2}+{\rm i}\epsilon)\, 
f^{2mh}\, \Big]
\nonumber \\
& &
{}-\frac{1}{2} \left( s-m_3^2-m_4^2+2 \frac{m_3^2 m_4^2}{t}
\right){\cal I}_3(m_3^2,m_4^2,s)
 \Bigg\}
.\label{f75}
\end{eqnarray}
\vspace{0.5cm}
\underline {The opposite ("easy") two$-$mass 
scalar box integral} 
\begin{eqnarray}
\lefteqn{I^{2me}_{4}(D=6;s,t;m_2^2,m_4^2) =
\frac{{\rm i}}{(4\pi)^{2}}\,\frac{1}{4 \pi \mu^2}\,
}\nonumber \\
& &{}\times h^{2me}\Bigg\{ \, \mbox{Li}_{2}
\Big[\, 1-(s+{\rm i}\epsilon)\, f^{2me}\, \Big]
+\mbox{Li}_{2}\Big[\,
1-(t+{\rm i}\epsilon)\, f^{2me}\, \Big] \nonumber \\
& &
{}
-\mbox{Li}_{2}\Big[\, 1-(m_{2}^{2}+{\rm i}\epsilon)\, 
f^{2me}\, \Big]
-\mbox{Li}_{2}\Big[\,
1-(m_{4}^{2}+{\rm i}\epsilon)\, f^{2me}\, \Big]\, \Bigg\}.
\label{f76}
\end{eqnarray}
\vspace{0.5cm}
\underline {The one$-$mass scalar box integral} 
\begin{eqnarray}
\lefteqn{I^{1m}_{4}(D=6; s,t;m_4^2)=
\frac{{\rm i}}{(4\pi)^{2}}\,\frac{1}{4 \pi \mu^2}\,
}\nonumber \\
& &{}\times h^{1m}\Bigg\{ \, \mbox{Li}_{2}
\Big[\, 1-(s+{\rm i}\epsilon)\, f^{1m}\, \Big]
+\mbox{Li}_{2}\Big[\,
1-(t+{\rm i}\epsilon)\, f^{1m}\, \Big] \nonumber \\
& &
{}
-\mbox{Li}_{2}\Big[\,
1-(m_{4}^{2}+{\rm i}\epsilon)\, f^{1m}\, \Big]
-\frac{{\pi}^2}{6}\, \Bigg\}.
\label{f77}
\end{eqnarray}
\vspace{0.5cm}
\underline {The zero$-$mass (massless) 
scalar box integral} 
\begin{eqnarray}
\lefteqn{I^{0m}_{4}(D=6; s,t) =
\frac{{\rm i}}{(4\pi)^{2}}\,
\frac{1}{4 \pi \mu^2}\,
}\nonumber \\
& &{}\times h^{0m}\Bigg\{ \, \mbox{Li}_{2}
\Big[\, 1-(s+{\rm i}\epsilon)\, f^{0m}\, \Big]
+\mbox{Li}_{2}\Big[\,
1-(t+{\rm i}\epsilon)\, f^{0m}\, \Big] 
-\frac{{\pi}^2}{3}\, \Bigg\},
\label{f78}
\end{eqnarray}
where
\begin{equation}
h^{K}=\left (-2\frac{\sqrt {{\rm det}(R_4^K)}}
{{~\rm det}(S_4^K)}\right ),\label{f79}
\end{equation}
and
\begin{equation}
\frac{\rm i}{(4\pi)^2}{\cal I}_3(a,b,c)
=I_3^{3m}(D=4;a,b,c)~.\label{f80}
\end{equation}
The functions appearing above are given by
\begin{eqnarray}
f^{3m}=f^{2me}&=&\frac{s+t-m_2^2-m_4^2}{st-m_2^2m_4^2}~,
\nonumber \\
f^{2mh}=f^{1m}&=&\frac{s+t-m_4^2}{st}~,\nonumber \\
f^{0m}&=&\frac{s+t}{st}~,\nonumber \\
h^{3m}&=&
\left (s+t-m_2^2-m_3^2-m_4^2
+m_3^2\frac{m_2^2t+m_4^2s-2m_2^2m_4^2}
{st-m_2^2m_4^2} \right)^{-1},
\nonumber \\
h^{2mh}&=&
\left (s+t-m_3^2-m_4^2+\frac{m_3^2m_4^2}{t} \right)^{-1},
\nonumber \\
h^{2me}&=&
\left (s+t-m_2^2-m_4^2 \right)^{-1},
\nonumber \\
h^{1m}&=&
\left (s+t-m_4^2 \right)^{-1},
\nonumber \\
h^{0m}&=&
\left (s+t \right)^{-1}. \nonumber
\end{eqnarray}

\section*{Appendix B}
\begin{figure}
\begin{center}
\includegraphics[width=8cm]{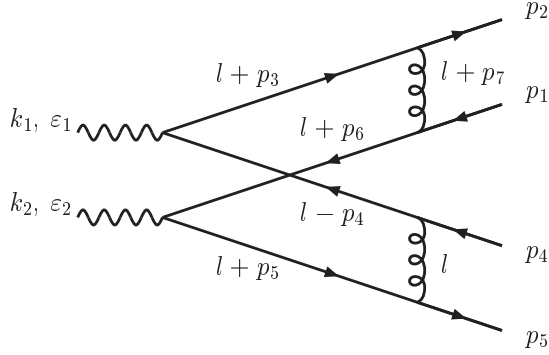}
\end{center}
\caption{\label{fig2}{\em One of the diagrams contributing to
the hard scattering amplitude 
of the process $\gamma \, 
\gamma \rightarrow 
\pi^+ \, \pi^-$ at NLO.}}
\end{figure}

As an illustration of the tensor decomposition and scalar 
reduction methods, we
evaluate an one-loop 6-point Feynman diagram shown in Fig. \ref{fig2}.

Note that, due to the kinematics which is bounded to three dimensional 
Minkowski subspace there are no four linearly independent four-vectors.
Consequently,
this diagram is of complexity of the 7-point one-loop
diagram with four dimensional external kinematics. We choose
this particular diagram because of the compactness of intermediate and
final expressions.

This is one (out of 462) diagram contributing to the NLO hard-scattering
amplitude for the exclusive process
$\gamma (k_1,\varepsilon_1)\, \gamma (k_2,\varepsilon_2)\rightarrow 
\pi^+(P_+) \, \pi^-(P_-)$,
(with both photons on-shell) at large momentum transfer.

In the $\gamma\,\gamma$ centre-of-mass frame, the 4-momenta of the incoming and
outgoing particles are
\begin{equation}
k_{1,\,2}={\sqrt s}/2\,(1,\mp \sin \theta_{c.m.},0,\pm \cos \theta_{c.m.}),
\quad
P_{\pm}={\sqrt s}/2\,(1,0,0,\pm 1),
\end{equation} 
while the polarization states of the photons are
\begin{equation}
\varepsilon_{1}^{\pm}=\varepsilon_{2}^{\mp}=
\mp 1/{\sqrt 2}\,(0,\cos \theta_{c.m.},
\pm i,\sin \theta_{c.m.}),
\end{equation}
where ${\sqrt s}$ is the total centre-of-mass energy of the $\gamma\,\gamma$
system (or the invariant mass of the $\pi^+\,\pi^-$ pair).

For example, taking $\theta_{c.m.}=\pi /2$ and assuming that the photons have
opposite helicities, the amplitude corresponding to the Feynman diagram of Fig.
\ref{fig2} is proportional to the integral 
\begin{equation}
{\cal I}=\frac{(\mu ^2)^{-\eir}}{2} \int \frac{d^{4+2 \eir} l}{(4 \pi)^{4+2 \eir}}
\frac{
{\rm Tr}\left[\gamma_{\mu} \gamma_{5} \Sl{P}_{+} 
\gamma^{\mu} (\Sl{l}+\Sl{p}_3) \Sl{\varepsilon}_1 
(\Sl{l}-\Sl{p}_4)\gamma_{\nu} \gamma_{5} \Sl{P}_{-} \gamma^{\nu}
(\Sl{l}+\Sl{p}_5)
\Sl{\varepsilon}_2 (\Sl{l}+\Sl{p}_6) 
\right]}{ l^2 (l+p_3)^2 (l-p_4)^2
(l+p_5)^2 (l+p_6)^2
(l+p_7)^2},\label{ex1}
\end{equation}
with the momenta $p_i$ $(i=1,\ldots,7)$ 
\begin{eqnarray}
& &{}p_1=\x\,P_{+},\,\,\,\,p_2=x\,P_{+},\,\,\,\,p_3=k_1-\y\,P_{-},\,\,
\,\,p_4=\y\,P_{-},\nn \\
& &{}p_5=y\,P_{-},\,\,\,\,p_6=y\,P_{-}-k_2,\,\,\,\,
p_7=\x\,P_{+}+y\,P_{-}-k_2.  
\end{eqnarray}
The quantities $x$ and $\x \equiv 1-x$ ($y$ and $\y \equiv 1-y$) 
are the fractions of the momentum
$P_+$ ($P_-$) shared between the quark and the antiquark in the $\pi^+$ 
($\pi^-$). 

With the aim of regularizing the IR divergences, the dimension of 
the integral is taken to be $D=4+2 \eir$.

The integral ${\cal I}$ is composed of one-loop 6-point tensor integrals 
of rank 
0, 1, 2, 3 and 4.
Performing the tensor decomposition and evaluating the
trace, we obtain the integral ${\cal I}$ in the form
\begin{eqnarray}
{\cal I}&=&-2{\left( 1 + \eir  \right) }^2\,{\Big [} 
    \,24\,s^3 \,x\,\x \,y\,\y\,
     {\left( 4\,\pi \,{\mu }^2 \right) }^4\,
     I_0^6(12 + 2\,\eir ,\{ 1,1,1,1,1,5\} )\nn \\  
& &{}+6\,s^3\,\y \,y\,
     {\left( 4\,\pi \,{\mu }^2 \right) }^4\,
     I_0^6(12 + 2\,\eir ,\{ 1,4,1,1,2,1\} ) \nn \\
& &{} + 
    2\,s^3\,\y \,\left( y - \y \right) \,
     {\left( 4\,\pi \,{\mu }^2 \right) }^4\,
     I_0^6(12 + 2\,\eir ,\{ 1,3,2,1,2,1\} )\nn \\
& &{} + 
    8\,s^3 \,y\,\y\,
     {\left( 4\,\pi \,{\mu }^2 \right) }^4\,
     I_0^6(12 + 2\,\eir ,\{ 1,3,1,1,3,1\} )\nn \\
& &{} + 
    2\,s^3\,x\,y\,\y\,
     {\left( 4\,\pi \,{\mu }^2 \right) }^4\,
     I_0^6(12 + 2\,\eir ,\{ 1,2,2,2,1,2\} )\nn \\
& &{}+ 
    s^3\,\y\,\left( y - \y \right) \,
     {\left( 4\,\pi \,{\mu }^2 \right) }^3\,
     I_0^6(10 + 2\,\eir ,\{ 1,2,2,1,2,1\} ) \nn \\
& &{} +s^2 \,y\,\y\,\left( 1 + \eir  \right) \,
     \left( 4\,\pi \,{\mu }^2\right )\,I_0^6(6 + 2\,\eir ,
      \{ 1,1,1,1,1,1\} )\nn \\
& &{} + s\,\left( 1 + \eir  \right) \,
     \left( 2 + \eir  \right) \,
     {\left( 4\,\pi \,{\mu }^2 \right) }^2\,
     I_0^6(8 + 2\,\eir ,\{ 1,1,1,1,1,1\} )\nn \\
& &{}+ \ldots \mbox{75 similar terms} \,{\Big ]}.\label{td}
\end{eqnarray}
Next, performing the scalar reduction using the method described in the 
paper, we
arrive at the following expression for the integral written in terms of the
basic integrals:
\begin{eqnarray}
\lefteqn{{\cal I}=8\,(1+\eir)^2 \left \{ (4\,\pi \,{\mu }^2)\left [
\frac{\eir}{\x} \,I_{4}^{1m}( 6+2\,\eir ;- s/2,-s\,\y/2;-s\,y/2)\right. \right.}
\nn \\
& &{}+\frac{1+\eir}{\x} \,I_{4}^{1m}( 6+2\,\eir ;- s\,y/2,-s/2;-s\,\y/2)\nn \\
& &{}+\left (1+\eir\left (1-\frac{x}{\x}\right )\right ) 
\,I_{4}^{1m}( 6+2\,\eir ;- s\,x/2,-s\,\y/2;-s\,(\x\,\y+x\,y)/2)\nn \\
& &{}\left.+\left( -\frac{x}{\x}+\eir\left (1-\frac{x}{\x}\right )\right)
\,I_{4}^{2me}( 6+2\,\eir ;- s\,\x/2,-s\,\y/2;-s/2,-s\,(\x\,\y+x\,y)/2)\right ]
\nn \\
& &{}+\frac{1}{s}\left [\frac{1}{(\x-x)\,y}\left( \frac{2\,\x}{\eir\,(\x-x)}+2-\frac{\x}{x}
\right)\,I_2(4+2\,\eir ;-s\,\x/2)\right.\nn \\
& &{}+\frac{1}{(x-\x)\,\y}\left( \frac{2\,x}{\eir\,(x-\x)}+2-\frac{x}{\x}
\right)\,I_2(4+2\,\eir ;-s\,x/2)\nn \\
& &{}+\frac{1}{(\y-y)\,x}\left( \frac{2\,\y}{\eir\,(\y-y)}+2-\frac{\y}{y}
\right)\,I_2(4+2\,\eir ;-s\,\y/2)\nn \\
& &{}+\frac{1}{(y-\y)\,\x}\left( \frac{2\,y}{\eir\,(y-\y)}+2-\frac{y}{\y}
\right)\,I_2(4+2\,\eir ;-s\,y/2)\nn \\
& &{}+\left( \frac{(1-x\,\y-3\,y\,\x)(1-y\,\x-3\,x\,\y)}{x\,\x\,y\,\y\,
(x-\x)(y-\y)}+\frac{2 (\x\,\y+x\,y)(8\,x\,\x\,y\,\y-x\,\x-y\,\y)}
{\eir\,x\,\x\,y\,\y\,
(x-\x)^2(y-\y)^2}\right)\nn \\
& &{}\left.\left.\times\,I_2(4+2\,\eir
;-s\,(\x\,\y+x\,y)/2)+\frac{(\x\,\y+x\,y)}{x\,\x\,y\,\y}\,I_2(4+2\,\eir
;-s/2)\right ] \right \} .\label{fin}
\end{eqnarray}
Here, $I_2$ is the two-point scalar integral in $D=4+2\,\eir$ 
with $\nu_i=1$, while
$I_{4}^{1m}$ and $I_{4}^{2me}$ are box scalar integrals 
in $D=6+2\,\eir$. Analytic
expressions for these integrals are given in the Appendix A. 
Expanding Eq. (\ref{fin}) up to order ${\cal O}(\eir ^0)$, we finally get
\begin{eqnarray}
\lefteqn{{\cal I}= \frac{i}{(4\,\pi)^2}\frac{8}{s}\left \{-\frac{1}{x\,\y}{\rm
Li}_2 (\x -x)-\frac{1}{y\,\x}{\rm
Li}_2 (x -\x)-\frac{1}{y\,\x}{\rm
Li}_2 (\y -y)-\frac{1}{x\,\y}{\rm
Li}_2 (y -\y)\right.}\nn \\
& &{}+\frac{x\,\y+y\,\x}{x\,\x\,y\,\y}
{\rm
Li}_2 \left( -(x-\x)(y-\y)\right)
+\frac{\pi ^2}{6}\frac{x\,\y+y\,\x}{x\,\x\,y\,\y}-
\frac{\x\,\y+x\,y}{x\,\x\,y\,\y}\ln \!\left(\frac{s}{2\,\mu^2}
\right)\nn \\
& &{}+
\frac{ (\x-2\,x)}{x\,y\,(\x-x)}\,\ln \left(\frac{s\,\x}{2\,\mu^2}
\right)+\frac{ (x-2\,\x)}{\x\,\y\,(x-\x)}\,
\ln \left(\frac{s\,x}{2\,\mu^2}
\right) \nn \\
& &{}+\frac{ (\y-2\,y)}{x\,y\,(\y-y)}\,\ln \left(\frac{s\,\y}{2\,\mu^2}
\right)+\frac{ (y-2\,\y)}{\x\,\y\,(y-\y)}\,
\ln \left(\frac{s\,y}{2\,\mu^2}
\right)\nn \\
& &{}-\frac{x}{\y\,(x-\x)^2}\,\ln ^2\left(\frac{s\,x}{2\,\mu^2}
\right)-\frac{\x}{y\,(x-\x)^2}\,\ln ^2\left(\frac{s\,\x}{2\,\mu^2}
\right)\nn \\
& &{}-\frac{y}{\x\,(y-\y)^2}\ln ^2\!\left(\frac{s\,y}{2\,\mu^2}
\right)-\frac{\y}{x\,(y-\y)^2}\ln ^2\!\left(\frac{s\,\y}{2\,\mu^2}
\right)\nn \\
& &{}-\frac{(1-x\,\y-3\,y\,\x)(1-y\,\x-3\,x\,\y)}{x\,\x\,y\,\y\,
(x-\x)(y-\y)}\,\ln \left(\frac{s\,(\x\,\y+x\,y)}{2\,\mu^2}
\right)\nn \\
& &{}-\frac{(\x\,\y+x\,y)(8\,x\,\x\,y\,\y-x\,\x-y\,\y)}{x\,\x\,y\,\y\,
(x-\x)^2(y-\y)^2}\,\ln ^2 \left(\frac{s\,(\x\,\y+x\,y)}{2\,\mu^2}
\right)\nn \\
& &-\frac{2}{\hat{\eir}}\left [ \frac{ x}{\y\,(x-\x)^2}\,
\ln \left(\frac{s\,x}{2\,\mu^2}
\right)+\frac{\x}{y\,(x-\x)^2}\,
\ln \left(\frac{s\,\x}{2\,\mu^2}
\right)\right.\nn \\
& &{}+\frac{ y}{\x\,(y-\y)^2}\,
\ln \left(\frac{s\,y}{2\,\mu^2}
\right)+\frac{\y}{x\,(y-\y)^2}\,
\ln \left(\frac{s\,\y}{2\,\mu^2}
\right)\nn \\
& &{}\left.\left.+\frac{(\x\,\y+x\,y)(8\,x\,\x\,y\,\y-x\,\x-y\,\y)}{x\,\x\,y\,\y\,
(x-\x)^2(y-\y)^2}\,\ln  \left(\frac{s\,(\x\,\y+x\,y)}{2\,\mu^2}
\right)\right ] \right \} ,\label{fin1}
\end{eqnarray}
where $1/\hat{\eir}=1/\eir+\gamma-\ln (4\,\pi)$.


\begin{thebibliography}{99}

\bibitem{old} G. Passarino and  M. Veltman, 
Nucl.\ Phys.\ B {\bf 160} (1979) 151;
G. J. van Oldenborgh and J. A. M. Vermaseren, 
Z.\ Phys.\ C {\bf 46} (1990) 425; 
W. L. van Neerven and J. A. M. Vermaseren, 
Phys.\ Lett.\ B {\bf 137} (1984) 241.


\bibitem{davy}
A.~I.~Davydychev,
Phys.\ Lett.\ B {\bf 263} (1991) 107.


\bibitem{dixon}
Z.~Bern, L.~J.~Dixon and D.~A.~Kosower,
Phys.\ Lett.\ B {\bf 302} (1993) 299
[Erratum-ibid.\ B {\bf 318} (1993) 649]
[arXiv:hep-ph/9212308].


\bibitem{bern}
Z.~Bern, L.~J.~Dixon and D.~A.~Kosower,
Nucl.\ Phys.\ B {\bf 412} (1994) 751
[arXiv:hep-ph/9306240].


\bibitem{tarasov}
O.~V.~Tarasov,
Phys.\ Rev.\ D {\bf 54} (1996) 6479
[arXiv:hep-th/9606018].
\bibitem{tarasov1}
J.~Fleischer, F.~Jegerlehner and O.~V.~Tarasov,
Nucl.\ Phys.\ B {\bf 566} (2000) 423
[arXiv:hep-ph/9907327].


\bibitem{binoth}
T.~Binoth, J.~P.~Guillet and G.~Heinrich,
Nucl.\ Phys.\ B {\bf 572} (2000) 361
[arXiv:hep-ph/9911342];
G.~Heinrich and T.~Binoth,
Nucl.\ Phys.\ Proc.\ Suppl.\  {\bf 89} (2000) 246
[arXiv:hep-ph/0005324].

\bibitem{binoth1}
T.~Binoth, J.~P.~Guillet, G.~Heinrich and C.~Schubert,
Nucl.\ Phys.\ B {\bf 615} (2001) 385
[arXiv:hep-ph/0106243].


\bibitem{glover}
J.~M.~Campbell, E.~W.~Glover and D.~J.~Miller,
Nucl.\ Phys.\ B {\bf 498} (1997) 397
[arXiv:hep-ph/9612413].

\bibitem{denner}
A.~Denner and S.~Dittmaier,
Nucl.\ Phys.\ B {\bf 658} (2003) 175
[arXiv:hep-ph/0212259].

\bibitem{mi1}
G.~Duplan\v ci\' c and B.~Ni\v zi\' c,
Eur.\ Phys.\ J.\ C {\bf 20} (2001) 357
[arXiv:hep-ph/0006249].

\bibitem{mi2}
G.~Duplan\v ci\' c and B.~Ni\v zi\' c,
Eur.\ Phys.\ J.\ C {\bf 24} (2002) 385
[arXiv:hep-ph/0201306].

\bibitem{davy1}
A.~I.~Davydychev,
J.\ Phys.\ A {\bf 25} (1992) 5587.

\bibitem{ibp}
K.~G.~Chetyrkin and F.~V.~Tkachov,
Nucl.\ Phys.\ B {\bf 192} (1981) 159;
F.~V.~Tkachov,
Phys.\ Lett.\ B {\bf 100} (1981) 65.

\bibitem{tarnew}
F.~Jegerlehner and O.~Tarasov,
Nucl.\ Phys.\ Proc.\ Suppl. {\bf 116} (2003) 83
[arXiv:hep-ph/0212004].

\bibitem{three}
A.~T.~Suzuki, E.~S.~Santos and A.~G.~Schmidt,
Eur.\ Phys.\ J.\ C {\bf 26} (2002) 125
[arXiv:hep-th/0205158].

\bibitem{box}
C.~Anastasiou, E.~W.~Glover and C.~Oleari,
Nucl.\ Phys.\ B {\bf 565} (2000) 445
[arXiv:hep-ph/9907523].
A.~T.~Suzuki, E.~S.~Santos and A.~G.~Schmidt,
arXiv:hep-ph/0210083.





\end{thebibliography}
\end{document}